\documentstyle[11pt]{article}
\newcommand{\be}{\begin{eqnarray}}

\newcommand{\ee}{\end{eqnarray}}

\title{
	\begin{flushright}
	{\normalsize TPI--MINN--54 \\
	NUC--MINN--93/30-T \\
        UMN--TH--1226/93 \\
	November 1993 \\}
	\end{flushright}
\bf Scattering in the Presence of Electroweak
Phase Transition Bubble Walls }
\author{
	Alejandro Ayala, Jamal Jalilian-Marian and
	Larry McLerran \\
	{\small\it School of Physics and Astronomy,
 	University of Minnesota, Minneapolis, MN 55455} \\
	\and
	Axel P. Vischer \\
	{\small\it Oregon State University, Corvallis, Oregon 97331} \\
	{\small\it School of Physics and Astronomy,
	University of Minnesota, Minneapolis, MN 55455} }
\date{}

\begin{document}

\maketitle

\begin{center}
{\bf Abstract}\\
\end{center}

We investigate the motion of fermions in the presence of an electro\-weak
phase transition bubble wall.
We derive and solve the Dirac equation for such fer\-mions, and compute
the transmission and reflection coefficients for fermions traveling
from the symmetric to the asymmetric phases separated by the domain wall.

\bigskip
\noindent PACS number(s): 12.15.Ji, 98.80.Cq
\vfill \eject

\section{Introduction}

There has been much recent work which suggests that the baryon
asymmetry of the universe might be produced at the electroweak
phase transition~\cite{review}. If the equations of motion of
electroweak theory are quantized, then anomalies arise
which are responsible for the
Chern--Simons --or baryon number changing-- currents~\cite{thooft}.
The rate of baryon number
violation is rather small for low temperatures but it becomes large at a
temperature scale on the order of the electroweak scale of about
100 GeV~{\cite{manton} - \cite{arnold}}.
Since baryon number is violated, a net baryon number will be generated
if there is a mechanism that biases the rate at the electroweak
phase transition in such a way, that
as the baryon number violation shuts off in the low temperature phase,
an asymmetry remains frozen into the system.
This can happen in practice if the electroweak phase transition is of
first order, and if there is sufficient CP violation
at the temperature of the phase transition.

It is expected that
the electroweak phase transition is of first order~{\cite{linde}
-\cite{linde1}}. In a first order phase transition, the conversion from
one phase to another occurs through nucleation of the true phase in the
false phase.  This happens when the system is either supercooled or
superheated.  The bubbles of the true phase
expand rapidly eating up the region
of the false phase.  For the electroweak phase transition,
this true phase eventually fills the entire volume with no
intermediate mixed phase~\cite{mag2}.  At the bubble surface,
there is a thin wall of microscopic dimensions which separates the
phases.  It is at this bubble wall that matter is
strongly out of equilibrium, and here the baryon asymmetry
is generated~{\cite{shaposhnikov} - \cite{farrar}}.

To quantitatively understand the generation of the baryon asymmetry
in the bubble wall, the effect of the bubble wall on the
propagation of fermions should be understood.
Fermions passing through the bubble or domain
wall acquire mass, generated by the Yukawa coupling, which is proportional to
the finite temperature vaccuum expectation value (vev) of the Higgs field.
This vev is determined from the equations of motion of the finite temperature
effective action of the bubble.

To simplify the problem, we will work in the approximation where
the energy density of the two phases are degenerate.  In practice this
is a good approximation, since in cosmology, the universe is expanding
so slowly that the nucleation always begins at a temperature
where the amount of supercooling is very small,
and the energies of the two phases are degenerate.
In this approximation, there is a one dimensional kink solution which separates
the phases, and the bubble wall is a domain wall which separates
the domains of different energy.  This kink wall can propagate
at any velocity.  In practice, the wall velocity is determined
by a complicated analysis which involves computing the effects
of dissipative processes, and is in the range 0.1 -- 0.9 of the speed of
light~{\cite{linde1}, \cite{turok} -- \cite{khlebnikov}}.
In our analysis, we will consider fermion propagation in
the presence of the wall at rest.  The case for a moving wall can be
determined by Lorentz boosting to the moving wall frame.

A plot of the domain wall is shown in Fig. 1.  At $x = -\infty$,
the system is in the symmetric phase, that is outside the bubble.
At $x = +\infty$, the system is in the symmetry broken phase, that
is, inside the bubble.  The approximation of the bubble as a planar
interface should be valid for bubbles which are large compared to
a microscopic size scale.  This is true for most of the evolution of
bubbles produced in the electroweak phase transition~\cite{mag2}.

The outline of this paper is as follows:
in Section 2 we describe how to obtain
the equation of motion for fermions in the presence of a domain wall
in the minimal standard model.
We obtain the Dirac equation with an effective mass proportional
to the vev of the Higgs field.
We solve this equation in Section 3,
and compute the fermion wave function we obtain transmission and reflection
coefficients. In Section 4, we present the normalized solutions of the
Dirac equation.
We finish by summarizing our results in section 5.

\section{The Motion of Fermions in the Presence of Domain Walls}

In this section we show how to obtain simplified classical equations of
motion for fermions in the standard model scattering from and interacting with
the electroweak domain walls. We use the classical mean-field approximation
in which the bosonic field operators are replaced by their classical
expectation values.

We will briefly review the notation which we will use throughout the paper.
The Lagrangian in the standard model is invariant under $ SU(2) \times U(1)$
transformations and the fields are therefore eigenstates of weak isospin
and hypercharge Y.
The $ SU(2) \times U(1) $ invariant vacuum state of the Lagrangian exists
only in a high temperature phase above the electroweak phase transition.
The symmetry is broken spontaneously once we cool the system below the
transition temperature $T_{c}$.

The Lagrangian is
\be
{\cal L} = {\cal L}_{gauge field} + {\cal L}_{Higgs} + {\cal L}_{fermion} +
	{\cal L}_{Yukawa} + {\cal L}_{gf} .
\ee

The gauge field, fermion, scalar kinetic energy
and gauge fixing terms are all written in the
standard way ~\cite{stm}.  For the potential energy of the Higgs field, we take
\be
	V(\Phi) = \lambda ( \Phi^\dagger \Phi - v^2/2)^2
\ee
so that the vacuum expectation value of the Higgs field is
\be
	<\Phi > = {v \over {\sqrt{2}}}.
\ee

To describe kink solutions and a first order phase transition, one has
to include the modifications of the effective potential for
the Higgs field due to interactions with the heat bath.
Including effects of both one-loop diagrams and the sum of ring diagrams,
the effective potential can be written to a good approximation as
\be
V_{eff} (\phi,T) = \frac{\gamma}{2} (T^{2} - T^{2}_{c} ) {\phi}^{2}
	- \delta T {\phi}^{3} + \frac{\lambda}{4} {\phi}^{4}
\ee
where $\phi = \sqrt{2} (\Phi^\dagger \Phi )^{1/2} $.  The dimensionless
parameters $\gamma$ and $\delta$ are approximately given by
\be
\gamma \sim
	\frac{5}{16} \mbox{g}^{2} \nonumber \\
\delta \sim
        \frac{1}{16 \pi} \mbox{g}^{3}
\ee
when $M_{W} \sim M_{Z} \sim M_{t} $, where these are the masses of
the W--bosons, the Z--boson and the top respectively.
$\delta$ is small and is responsible for generating the
first order phase transition.  At some value of $T$, there are two
degenerate minima of this potential and a kink solution exists.

Of course the above evaluation of the effective potential may
not be such a good approximation for realistic values of the
Higgs mass.  If the effective potential is greatly modified then
the kink solution, estimates for bubble wall velocity and the
considerations we present will all have to be modified.  Nevertheless,
in the words of an old U. S.  television show, {\it it is better to light one
little candle than to curse the darkness}.

As the scalar field evolves through the kink solution,
no current is generated for
the vector field, so that no expectation value of the vector fields
is generated. We are therefore justified in
truncating the system to only the Higgs field degrees of freedom.
The equation of motion is just the classical Higgs field equation
of motion with zero background vector field.

To get the effective potential for the Higgs field into more transparent
form~\cite{turok},
we introduce the dimensionless temperature
$ \zeta = \frac{ \lambda \gamma }{{\delta}^{2}} (1- \left(
\frac{T_{c}}{T} \right)^{2} )$
and the dimensionless field strength $ g = \frac{\lambda}
{\delta T } \phi $ and obtain
\be
V_{eff} (g) = \delta T ( \frac{\delta T}{\lambda} )^{3}
	( \frac{\zeta}{2} g^{2} - g^{3} + \frac{1}{4} g^{4} )
\ee
The potential $ V_{eff} (g) $ is plotted in Figure 2 for different
values of $ \zeta $.
For large positive values of $ \zeta $ we are in the high temperature phase
with the vev of the Higgs field being zero. Decreasing $ \zeta $ we develop
at $ \zeta = 2.25 $ a second relative minimum which for $ \zeta = 2 $
becomes degenerate with the first one. For smaller values of $ \zeta $,
the high temperature phase at $ g = 0 $ becomes unstable with respect to the
new absolute minimum at $ g \neq 0 $ , the first order phase transition sets
in and bubbles of the broken phase start to nucleate.
If we supercool to $ \zeta = 0 $ the system spinodally decomposes.

Since the expansion rate of the universe $ ( \approx 10^{-11} s ) $
is rather slow compared to the
electroweak timescale $ ( \approx 10^{-26} s ) $ we cannot strongly
supercool and the phase transition will roughly proceed at $ \zeta \approx 2 $
where the minima of the effective potential are degenerate. The steady state
solution describing the domain wall is thus given approximately
by the solution of the Higgs field $g$ for a transition between the
degenerate minima at temperature $ \zeta = 2 $.

The equation of motion for the kink is mathematically
identical to treating the amplitude of the Higgs field $g$
as a spatial coordinate for the inverted potential
of the previous equation and taking the spatial coordinate of the kink as the
time variable.
The mechanical analog to this system is a frictionless ball rolling from
the top of one hill through a valley to the top of another
hill.  For degenerate minima the tops of the two hills have the
same height and there exists a solution where the ball starts at the top of
one hill from rest and ends on the top of the other hill again at rest.
Such a solution is the kink. Energy conservation for the kink reads
\be
E = \frac{1}{2} ( \frac{dg}{dr} )^{2} - V_{eff} (g(r))
\ee
which for a transition between degenerate minima $ (E = 0) $ and a kink
bubble with dimensionless position $ x = \frac{\delta}{\sqrt{2 \lambda}} T r $
can be rewritten as
\be
\frac{dg}{dx} = - g(g-2) .
\ee

We can integrate the above equation
implementing the boundary condition that at
the center of the nucleated bubble $ ( x \rightarrow + \infty ) $
the system is in the broken phase with a finite vev while outside the domain
wall
$ ( x \rightarrow - \infty ) $ it approaches a zero vev in the high
temperature phase, treating the domain wall as effectively
flat and separating two semi--infinite domains. We obtain
\be
g = 1 + \tanh (x).
\ee
This solution, commonly called the 'kink', is depicted in figure 1.
It represents the classical, steady state, finite temperature
solution to the equations of motion.

Given a background scalar field, we can solve the Dirac equation in
the presence of this field.  Rewriting the coordinates in terms of
ordinary dimensionful length scales, the Dirac equation is
\be
	\left( \not \! p -{{\delta T} \over {\sqrt{2\lambda}}} \xi
 g(x) \right) \Psi(x) = 0    \label{eq:dirac}
\ee
where $\Psi$ is any fermion field, and $\xi$ is twice the ratio of
quark to Higgs mass at zero temperature.  The solution to this
equation describes the motion of a fermion in the presence of a kink.
To this order of approximation for the standard model, CP violation plays
no essential role in this equation.

\section{Solving the Dirac equation}

In this section we study the solutions to the Dirac equation~(\ref{eq:dirac}).
We start by
expli\-cit\-ly showing how to solve this equation analytically for the
fermion wave function. The second part
is devoted to deriving the transmission and reflection coefficients from the
wave function.

\subsection{The wave function}

We will solve the Dirac equation by looking for eigenstates of
momentum in the plane of the wall, $p_t$, that is the momentum
perpendicular to a normal vector on the surface of the wall.
The Dirac equation~(\ref{eq:dirac}) is then a differential equation in one
dimension
and the only relevant variable ($r$) is the one along the direction normal to
the
domain wall. The degrees of freedom in the plane of the
domain wall effectively decouple. This can be used by substituting
the ansatz
\be
\Psi (r,x_{t},t)=(\not \! p + \frac{\delta T}{\sqrt{2 \lambda}} \xi g(r))
                 e^{\pm i (p_{t}x_{t}-Et) }\Phi (r) \label{eq:ansatz}
\ee
into equation~(\ref{eq:dirac}), finding
\be
(p^{2} - ( \frac{\delta T}{\sqrt{2 \lambda}} \xi g(r) )^{2}
	+ [ \not \! p , \frac{\delta T}{\sqrt{2 \lambda}} \xi g(r) ] )
	\Phi (r) = 0 \label{eq:Dirac 2}.
\ee
The $+$ and $-$ signs in equation~(\ref{eq:ansatz})
correspond to the positive and negative
energy solutions respectively.

The commutator in equation~(\ref{eq:Dirac 2}) picks only a contribution from
the
direction normal to the wall and therefore it is proportional to just one of
the
$ \gamma \! $ --matrices and we choose this to be $ \gamma_{3}$.
We will use the ordinary representation for gamma matrices where
\be
	\gamma^{i} = \left( \begin{array}{clcr} 0 & \sigma^{i} \\
                                  -\sigma^{i} & 0 \end{array} \right)
\label{eq:gammas}
\ee
The eignestates of $\gamma_{3} $ are
\be
	\gamma_{3} {\bf u}^{s}_\pm = \pm i {\bf u}^{s}_\pm \label{eq:u's}
\ee
where $s = 1,2$ and ${\bf u}^{s}_\pm $ are
\be
	{\bf u}^{1}_\pm = \left( \begin{array}{clcr} 1 \\
                                 0 \\
                             \pm i \\
                                 0 \end{array} \right)
\mbox{ and }
	{\bf u}^{2}_\pm = \left( \begin{array}{clcr} 0 \\
                                 1 \\
                                 0 \\
                             \mp i \end{array} \right) \label{eq:spinors}
\ee

Recalling that the dimensionless
radius $x $ is  $ x = \frac{\delta}{\sqrt{2 \lambda}} T r $, we obtain
from~(\ref{eq:Dirac 2}),
in dimensionless variables
\be
\{ \frac{d^{2}}{dx^{2}} +i\gamma_{3}\xi \frac{d}{dx} g(x) -\xi ^{2}g^{2}(x) +
\epsilon ^{2}\}\Phi (x)=0 \label{eq:igamma}
\ee

The dimensionless energy parameter $\epsilon$ is given in terms of the energy
E and the momentum parallel to the bubble wall $p_{t}$ by
\be
\epsilon =(\sqrt{2\lambda}/\delta T)\sqrt{E^{2}- p^{2}_{t}} \label{eq:epsilon}.
\ee

We write the spinor $\Phi (x)$ in the basis~(\ref{eq:spinors})
\be
\Phi (x) = \sum \phi_{\pm }^{s}(x) {\bf u}_{\pm }^{s} \label{eq:combi}
\ee
The spinors~(\ref{eq:spinors}) are linearly independent and thus,
writing~(\ref{eq:combi})
into equation~(\ref{eq:igamma}) we get two differential equations
for the functions $\phi_{\pm }^{s}$,
\be
\{ \frac{d^{2}}{dx^{2}} \mp \xi \frac{d}{dx} g(x) -\xi ^{2}g^{2}(x) +
\epsilon ^{2}\}\phi_{\pm } (x) = 0 \label{eq:components}.
\ee

To further simplify equations~(\ref{eq:components}), we factorize the
singularities. With the change of variable $z=\frac{1}{2}(1-\tanh (x))$ we
substitute
\be
\phi_{\pm }(x) = z^{\alpha }(1-z)^{\beta } \chi_{\pm }(z) \label{eq:sing}
\ee
and examine the behavior of the resulting differential equations near the
singular points
$z=0$ and $z=1$. Assuming that the functions $\chi_{\pm }$ are slowly varying
near the
singularities, this yields two algebraic conditions from which we get
\be
\alpha & = & +\frac{i}{2}\sqrt{\epsilon^{2}- 4\xi^{2}} \nonumber \\
\beta  & = & +\frac{i}{2}\epsilon \label{eq:alphabeta}
\ee
(the choice of signs for $\alpha $ and $\beta $
is discussed later). With the values of $\alpha $ and $\beta $ at hand, the
differential
equation for the functions $\chi_{\pm }$ is the hypergeometric differential
equation
\be
\{ z(1-z) \frac{d^{2}}{dx^{2}} + (c-(1+a_{\mp}+b_{\mp})z) \frac{d}{dx} -
a_{\mp}b_{\mp} \}
 \chi_{\pm}
	= 0   \label{eq:hyper}
\ee
with the parameters $ a_{\mp}, b_{\mp}$ and c given by
\begin{eqnarray}
 a_{\mp} & = & \alpha +\beta + \frac{1}{2} - \left| \xi \mp \frac{1}{2} \right|
             \nonumber  \\
 b_{\mp} & = & \alpha +\beta + \frac{1}{2} + \left| \xi \mp \frac{1}{2} \right|
             \nonumber \\
 c       & = & 2\alpha +1 \label{eq:parameters} .
\end{eqnarray}
Each of the equations~(\ref{eq:hyper}) has two independent
solutions~\cite{grad}
\be
\chi ^{(- \alpha)}_{\pm}(z) & = & \ _{2}F_{1}(a_{\mp},b_{\mp},c;z) \nonumber \\
\chi ^{(+ \alpha)}_{\pm}(z) & = & z^{1-c}\
_{2}F_{1}(a_{\mp}+1-c,b_{\mp}+1-c,2-c;z)
	\label{eq:chi}
\ee
The hypergeometric equations are expressed here as expansions around $z = 0$.
The superscripts $(\pm \alpha)$ are explained below.

The general solutions to equations~(\ref{eq:components}) are
\be
\phi_{\pm}^{s} = (A^{s})^{(- \alpha)}_{\pm} \phi^{(- \alpha)}_{\pm}
	+ (A^{s})^{(+ \alpha)}_{\pm} \phi_{\pm }^{(+ \alpha)}\label{eq:gensolphi}
\ee
with
\be
\phi _{\pm }^{(- \alpha)} & = & z^{\alpha}(1-z)^{\beta} \ _{2}F_{1}(a_{\mp
},b_{\mp },c;z) \nonumber \\
\phi _{\pm }^{(+ \alpha)} & = & z^{-\alpha}(1-z)^{\beta} \ _{2}F_{1}(a_{\mp
}+1-c,b_{\mp }+1-c,2-c;z)
\label{eq:solution phi}
\ee
where we have used~(\ref{eq:sing}). The superscripts $(\pm \alpha)$ indicate
the behavior of these solutions at $x \rightarrow + \infty$ or correspondingly
at $z = 0$ since
\be
z^{\pm \alpha} (1-z)^{\beta} & \stackrel
	{\mbox{\tiny $x \rightarrow + \infty $}}{\longrightarrow}
                                    & e^{\mp 2 \alpha x} \label{eq:lim1} \\
   z^{\alpha} (1-z)^{\pm \beta} & \stackrel
	{\mbox{\tiny $x \rightarrow - \infty $}}{\longrightarrow}
                                    & e^{\pm 2 \beta x} \label{eq:lim2}.
\ee
and $_{2}F_{1}(a,b,c;0) = 1$. The behavior of the solution in these limits
will be needed in the next section to implement boundary conditions.

We now proceed to find the fermion wave function.
We begin by rewriting equation~(\ref{eq:ansatz})
(in units of $\frac{\delta }{\sqrt{2 \lambda}} T$)
in a slightly different form
\be
\Psi (x) & = & \{\epsilon_{r} \tilde{\gamma} +i \gamma_{3} \frac{d}{dx} + \xi
g(x) \} \Phi (x)
\label{eq:ferm0}
\ee
where we have set
\be
\tilde{\gamma} & = & \gamma_{0} \cosh \theta - \gamma_{t} \sinh \theta
\nonumber \\
             E & = & \frac{\delta T}{\sqrt{2 \lambda}} \epsilon_{r} \cosh
\theta \nonumber \\
         p_{t} & = & \frac{\delta T}{\sqrt{2 \lambda}} \epsilon_{r} \sinh
\theta \label{eq:rotation},
\ee
$r=1,2$ and $\epsilon_{r}$ is given by
\be
\epsilon_{r} = \left\{ \begin{array}{c} \mbox{ } \epsilon , \mbox{ }r=1 \\
                                       -\epsilon , \mbox{ } r=2 \end{array}
\right.
\label{eq:posorneg}
\ee
with $\epsilon$ given by equation~(\ref{eq:epsilon}). The parameter $\theta $
is related
to the particle velocity in the direction parallel to the wall, $v_{t}$ by
$v_{t} = \tanh \theta $.

{}From now on, we set $p_{t} = 0 $ that is $\epsilon = E$
(in units of $\frac{\delta }{\sqrt{2 \lambda}} T$)
so that the fermion moves along
the direction normal to the domain wall. From~(\ref{eq:rotation})
we also see that
$\tilde{\gamma}$ becomes $\gamma_{0} $.
The general case for which the fermion has a non vanishing $p_{t}$
can be obtained by Lorentz boosting the solution along the direction of the
wall by
means of the transformation
\be
S(\theta) = \exp (- \gamma_{0} \gamma_{t} \theta ). \label{eq:trafo}
\ee

To finish this section, let us work with the $\epsilon $--independent
part of equation~(\ref{eq:ferm0}) and evaluate
\be
\{ +i \gamma_{3} \frac{d}{dx} + \xi g(x) \} \Phi (x) \label{eq:epsind}
\ee
using $\Phi $ given by~(\ref{eq:combi}) and~(\ref{eq:gensolphi}), with ${\bf
u}_{\pm}^{s}$ given by~(\ref{eq:spinors}),
the above equation becomes
\be
\sum \varphi_{\pm}^{s} {\bf u}_{\pm}^{s}
\label{eq:eps0}
\ee
with
\be
\varphi_{\pm}^{s} = (A_{\pm}^{s})^{(- \alpha)} \varphi^{(- \alpha)}_{\pm}
	+ (A_{\pm}^{s})^{(+ \alpha)} \varphi^{(+ \alpha)}_{\pm}
\label{eq:gensolvarphi}
\ee
where we have defined, using $\phi_{\pm }$ given by~(\ref{eq:gensolphi})
and~(\ref{eq:solution phi}),
\be
\varphi_{\pm } \equiv \{ \mp \frac{d}{dx} + \xi g(x) \} \phi_{\pm } (x)
\label{eq:epsindcomp}.
\ee

To construct the full wavefunction we  set
$\epsilon = \frac{\delta T}{\sqrt{2 \lambda }} E $ and
$\tilde{\gamma }=\gamma_{0}$ in~(\ref{eq:ferm0}) and boost then the so found
solution in the direction along the domain wall. We need to know therefore
how $\gamma_{0}$ operates on the spinors $\bf u$. In our representation
\be
\gamma_{0} = \left( \begin{array}{clcr} & I & 0 \\
                                        & 0 & -I \end{array} \right) \nonumber
,
\ee
and thus
\be
\gamma_{0} {\bf u}_{\pm } = {\bf u}_{\mp } \label{eq:gamma0}.
\ee

Recall that each of the four terms in equation~(\ref{eq:combi}) is by itself
a solution to equation~(\ref{eq:igamma}) and so for a given sign of the energy,
equation~(\ref{eq:ferm0}), with $\Phi (x)$ replaced by any of its four
components, represents four solutions from which only two should be
independent.
In appendix {\bf A} we show that $\varphi_{\pm}$ are related to $\phi_{\mp}$ by
\be
\varphi_{\pm}^{(- \alpha)} & = & 2(\xi \pm \alpha)\phi_{\mp}^{(- \alpha)}
\nonumber \\
\varphi_{\pm}^{(+ \alpha)}& = & 2(\xi \mp \alpha)\phi_{\mp}^{(+ \alpha)}
\label{eq:relationphis}.
\ee
We also notice that $\phi^{(+ \alpha)}_{\pm}(x)$ can be obtained from
$\phi^{(- \alpha)}_{\pm}(x)$ by exchanging $\alpha $ with $-\alpha $.
{}From equation~(\ref{eq:ferm0}) together with~(\ref{eq:gensolvarphi})
,~(\ref{eq:gamma0}) and~(\ref{eq:relationphis}), we get the four independent
solutions to the Dirac equation labeled by the indices $s$ and $r$
\be
\Psi_{r}^{s} & = & (A^{s})^{(- \alpha)}  ( \epsilon_{r} \phi_{+}^{(- \alpha)}
	{\bf u}_{-}^{s} +
                                  2(\xi + \alpha )\phi_{-}^{(- \alpha)}
	{\bf u}_{+}^{s} ) \nonumber \\
             & + & (A^{s})^{(+ \alpha)} ( \epsilon_{r} \phi_{+}^{(+ \alpha)}
	{\bf u}_{-}^{s} +
                                  2(\xi - \alpha )\phi_{-}^{(+ \alpha)}
	{\bf u}_{+}^{s} )\label{eq:totalPsi}
\ee
where the constants $(A^{s})^{(- \alpha)}$ and $(A^{s})^{(+ \alpha)}$
replace $(A^{s})^{(- \alpha)}_{\pm}$ and $(A^{s})^{(+ \alpha)}_{\pm}$
respectively and have to be fixed by normalization and the choice of initial
conditions. Again the general case for which the fermion has a non vanishing
$p_{t}$ can be obtained by Lorentz boosting~(\ref{eq:totalPsi}) along the
direction of the wall according to~(\ref{eq:trafo}).
Starting from the solution~(\ref{eq:totalPsi}) we will discuss in the next
section the scattering states of fermions traversing the domain wall.

\subsection{The scattering states}

At the electroweak phase transition the temperature is larger
than the typical particle masses such as that of the Z boson or light mass
quarks and although it can be of the order of the top quark mass,
most of the fermions will pass by the
domain wall with energies above its height.
To understand the interaction of fermions of such energies with the wall
we proceed to study the scattering states of the fermion wave function and to
derive reflection and transmission coefficients.

The two physical processes of interest here are either a fermion leaving the
bubble or a fermion falling into the bubble. We call the solutions
corresponding
to the boundary conditions of these processes the type $I$ and type $II$
solutions respectively. Both processes are depicted in Figure 3.

For the type $I$ solution we have an incident fermion from the right ($ x
\rightarrow + \infty $) with an energy parameter $\epsilon$ which is always
larger than the height of the wall ( $\epsilon^2 \geq 4 \xi^2$). At the wall
this fermion is scattered into an reflected wave going to the right and a
transmitted wave going to the left ($x \rightarrow - \infty$). Overall the
fermion is therefore represented by an incoming and reflected wave to the
right of the wall and by a transmitted wave to the left.

If a fermion is incident from the left (type $II$ solution) it can approach
the wall with energies above its height ($\epsilon^2 \geq 4 \xi^2$) or below
its height ($\epsilon^2 \leq 4 \xi^2$). In any case there will be again the
reflected wave running now to the left and the transmitted wave running to
the right. But for energies below the barrier ($\epsilon \leq 4 \xi^2$) the
transmitted wave has to tunnel through the wall leading to a decaying instead
of an oscillating wave.

In this section we start out by constructing the wavefunction corresponding to
the boundary conditions of the type $II$ solution and determine then the
transmission and reflection coefficients for this case. We show then that we
need to rewrite the hypergeometric functions as expansions around $z = 1$ to
obtain the wavefunction for the boundary conditions of type $I$. It turns out
that the transmission and reflection coefficients for this case are identical
to the ones obtained for type $II$.

For a given pair of indices $s$ and $r$, equation~(\ref{eq:totalPsi}) consists
of two terms with different asymptotic behaviors as $x \rightarrow +\infty $.
In this limit, according to equation~(\ref{eq:lim1}), the first term
proportional to $\phi^{(- \alpha)}_{\pm}$ behaves like $\exp(-2\alpha x)$
whereas the second term proportional to $\phi^{(+ \alpha)}_{\pm}$ behaves
like $\exp(2\alpha x)$. The boundary conditions appropriate to the description
of particles crossing the wall from the symmetric to the asymmetric phase are
as follows:
At $x=- \infty$, corresponding to $ z = 1 $ (outside the
bubble), we require the solutions~(\ref{eq:totalPsi})
to describe two plane waves,
one moving towards
(incoming wave $ {\Psi}^{\it inc} ) $ and the other away from the wall
(reflected wave $ {\Psi}^{\it ref} ) $. At $x=+ \infty$, corresponding to
$ z = 0 $ (inside the bubble), we impose that for energies such that
$\epsilon^{2} > 4 \xi^{2}$
there is only a single plane wave moving away from the wall (transmitted
wave $ {\Psi}^{\it trans} $),  while for $\epsilon^{2} \leq 4 \xi^{2}$ the
solution dies out exponentially.

The above conditions require that $(A^{s})^{(- \alpha)}=0$
in~(\ref{eq:totalPsi}) and the wave function for case $II$ looks therefore
like
\be
(\Psi ^{s}_{r})_{II} = A_{II}( \epsilon_{r} \phi_{+}^{(+ \alpha)} {\bf
u}_{-}^{s}
		+ 2(\xi - \alpha )\phi_{-}^{(+ \alpha)} {\bf u}_{+}^{s} )
\label{eq:completesolutionII}.
\ee
The normalization constant $A_{II}$ is determined in the next section and turns
out to be independent of the quantum numbers $s$ and $r$.

To compute the transmission and reflection coefficients, we need first to look
at the behavior of the solutions as $x \rightarrow \pm \infty$.

If we take the limit $ x \rightarrow + \infty$ of
equation~(\ref{eq:completesolutionII}) we find
\be
(\Psi_{r}^{s})_{II}  \stackrel{\mbox{\tiny $x \rightarrow + \infty
$}}{\longrightarrow}
(\Psi _{r}^{s})^{\it trans}_{II} = A_{II} (\epsilon_{r}{\bf u}_{-}^{s}
				+ 2( \xi -\alpha ){\bf u}_{+}^{s}) e^{2\alpha x}
\label{eq:PsitransII}
\ee

For $x \rightarrow - \infty $,
we have first to evaluate the hypergeometric functions in the second equation
of~(\ref{eq:solution phi}) at $z=1$. Since these functions are defined as
expansions around $z=0$ they are ill defined at $z=1$ and we need to use the
identity~\cite{grad}
\be
_{2}F_{1}(a,b;c;z) & = &  \frac{\Gamma (c) \Gamma (c-a-b)}{\Gamma (c-a)
	                  \Gamma(c-b)} {_{2}F_{1}}(a,b,a+b-c+1;1-z) \nonumber \\
                \, & + &  (1-z)^{c-a-b}
	                  \frac{\Gamma (c) \Gamma (a+b-c)}{\Gamma (a) \Gamma (b) }
\nonumber \\
	        \, & \times & {_{2}F_{1}} (c-a,c-b,c-a-b+1;1-z)
\label{eq:ident}.
\ee
therefore, using equation.~(\ref{eq:ident})
in~(\ref{eq:solution phi}) and considering
the limit $x \rightarrow - \infty $ we get from~(\ref{eq:completesolutionII})
after some algebra
\be
(\Psi_{r}^{s})_{II} \stackrel{\mbox{\tiny $x \rightarrow - \infty
$}}{\longrightarrow}
(\Psi_{r}^{s})^{\it inc}_{II} + (\Psi_{r}^{s})^{\it ref}_{II} \nonumber
\ee
\be
(\Psi_{r}^{s})^{\it inc}_{II} & = & \frac{A_{II} \Gamma (1 - 2\alpha)\Gamma
(-2\beta )}
	                    {\Gamma (-\alpha -\beta +\xi)\Gamma (-\alpha -\beta -\xi
)}
                            \left( \frac {\epsilon_{r}{\bf u}_{-}^{s}}{-\alpha
-\beta - \xi} +
                                   \frac {2(\xi - \alpha){\bf
u}_{+}^{s}}{-\alpha -\beta + \xi}
                            \right) e^{2 \beta x} \nonumber \\
(\Psi_{r}^{s})^{\it ref}_{II} & = & \frac{A_{II} \Gamma (1- 2\alpha)\Gamma (2
\beta )}
	                    {\Gamma (-\alpha +\beta +\xi)\Gamma (-\alpha +\beta -\xi
)}
	   	            \left( \frac {\epsilon_{r}{\bf u}_{-}^{s}}{-\alpha +\beta -
\xi} +
                                   \frac {2(\xi - \alpha){\bf
u}_{+}^{s}}{-\alpha +\beta + \xi}
                            \right) e^{-2 \beta x } \nonumber \\
\label{eq:inreII}.
\ee
We can check that the reflected wave can be obtained from the incident
by exchanging $ \beta $ with $ - \beta $.

To compute the reflection and transmission coefficients, we have to compute the
ratio
of the reflected and transmitted fluxes to the incoming one. It is thus
sufficient
to calculate the ratios of the normal components of the corresponding vector
currents.

The normal component of the currents associated with the plane waves that we
have found are
\be
j_{3} = {\overline{\Psi}} \gamma_{3} \Psi
\label{eq:current}
\ee
where $\Psi $ is any of $\Psi^{\it trans}, \Psi^{\it inc}$ or $\Psi^{\it ref}$.
{}From our solution, equation~(\ref{eq:completesolutionII}), the normal
components of the
incident, reflected and transmitted currents are for type $II$
\be
(j_{II}^{\it inc})_{3} & = & \; \: 4 |A_{II}|^{2}\epsilon \epsilon_{r}
                         \left| \frac{\Gamma (1 + 2\alpha )\Gamma (2\beta )}
                                     {\Gamma (1 +\alpha +\beta +\xi)\Gamma
(\alpha +\beta -\xi )}
                        \right| ^{2} \nonumber \\
(j_{II}^{\it ref})_{3} & = & \! \!-4 |A_{II}|^{2}\epsilon \epsilon_{r}
                         \left| \frac{\Gamma (1 + 2\alpha )\Gamma (2\beta )}
                                     {\Gamma (1 +\alpha -\beta +\xi)\Gamma
(\alpha -\beta -\xi )}
                        \right| ^{2} \nonumber \\
(j_{II}^{\it trans})_{3} & = & \; \: 8 |A_{II}|^{2}\epsilon_{r} |\alpha
|\nonumber \\
\label{eq:fnalcurrII}
\ee

The reflection and transmission
coefficients $R$ and $T$ are now the ratios of the
reflected and transmitted normal currents to the incident one, respectively,
projected along a unit vector normal to the domain wall.
Using equations~(\ref{eq:fnalcurrII})
and the identity~\cite{grad}
\be
\Gamma (x) \Gamma (-x) = \frac{-\pi}{x \sin (\pi x)},
\ee
we find after some simplification
\be
T  & = & \frac{-\sin 2\pi \alpha \ \sin 2\pi \beta }
         {\sin \pi (\xi -\alpha -\beta )\  \sin \pi (\xi +\alpha +\beta )}
\nonumber \\
R  & = & \frac{\sin \pi (\xi +\alpha -\beta )\  \sin \pi (\xi -\alpha +\beta )}
         {\sin \pi (\xi -\alpha -\beta )\  \sin \pi (\xi +\alpha +\beta )}
\label{eq:RT}.
\ee

{}From the above equations we see that for states with energy parameter
$\epsilon^{2} = 4 \xi^{2}$,
the transmission and reflection coefficients become $0$ and $1$.
For states with $\epsilon^{2} < 4\xi^{2}$,
$\alpha $ becomes negative and real. From equation~(\ref{eq:PsitransII}) we see
that
$ j^{\it trans}$ is identically zero
and therefore the transmission coefficient $T$ is also zero. Correspondingly,
the
reflection coefficient $R$ for fermions with $\epsilon^{2} < 4 \xi^{2}$ is one.

Both reflection and transmission coefficients are the same for positive
and negative energy solutions and thus for fermions and antifermions.
They are depicted in figure 4. In this and the following
figures we choose two representative values for the parameter $\xi$.
The heavier fermions are represented by $\xi = 4$ which corresponds to a ratio
of fermion to Higgs mass of 2. It is expected that the mass of the top quark is
lying in this mass region, so that we can consider the results for $\xi = 4$
to represent a top interacting with the domain wall. The other fermions, on
the other hand are much lighter than the Higgs. The corresponding value of
$\xi$ will be therefore very small. We choose a value of $\xi = 0.6$ to
represent these particles which corresponds to a ratio of fermion to Higgs
mass of 0.3.

We plot in figure 4 the transmission and reflection coefficients as a function
of a reduced energy parameter $y = \frac{\epsilon}{2 \xi}$. The top of the
barrier is then at $y=1$. We see that the interactions between the heavy
fermions and the wall quickly die out if we increase the energy parameter.
The light fermions on the other hand are still feeling the presence of the
wall high above the top of barrier.

The result for the reflection and transmission coefficient is unchanged if
we boost the solution along the direction of the wall to go to non--zero
$p_{t}$. Since the boost--operator defined
in~(\ref{eq:trafo}) commutes with $\gamma_{3}$
\be
[ \gamma_{3} , S(\theta) ] = 0 \nonumber
\ee
we see that the currents defined in~(\ref{eq:current}) is invariant under
$S$. This shows that also~(\ref{eq:RT}) is unchanged.

We proceed now to the solution of type $I$. Here the transmitted wave has to
behave like $e^{-2 \beta x}$ for $x \rightarrow - \infty$. But the solution
{}~(\ref{eq:totalPsi}) is not regular in this limit -- the hypergeometric
functions
diverge. To resolve this problem we have to rewrite~(\ref{eq:totalPsi}) in
terms
of hypergeometric functions defined as expansions around $z=1$.

There are two possible ways to obtain this modification of~(\ref{eq:totalPsi}).
One possibility is to restart from~(\ref{eq:chi}) and write the two independent
solutions of the hypergeometric equation as hypergeometric functions defined
as expansions around $z=1$. Equation~(\ref{eq:chi})
becomes then~(\cite{erdelyi})
\be
\chi ^{(+ \beta)}_{\pm}(z) & = &
	\ _{2}F_{1}(a_{\mp},b_{\mp},a_{\mp}+b_{mp}-c+1;1-z) \nonumber \\
\chi ^{(- \beta)}_{\pm}(z) & = & (1-z)^{c-a_{\mp}-b_{\mp}}
	\ _{2}F_{1}(c-a_{\mp},c-b_{\mp},c-a_{\mp}-b_{\mp}+1;1-z)
\label{eq:chinew}
\ee
and~(\ref{eq:solution phi}) is now
\be
\phi _{\pm }^{(+ \beta)} & = & z^{\alpha}(1-z)^{\beta}
	\ _{2}F_{1}(a_{\mp },b_{\mp },a_{\mp}+b_{\mp}-c+1;1-z) \nonumber \\
\phi _{\pm }^{(- \beta)} & = & z^{\alpha}(1-z)^{-\beta}
	\ _{2}F_{1}(c-a_{\mp },c-b_{\mp },c-a_{\mp}-b_{\mp}+1;1-z).
\label{eq:phinew}
\ee
Finally we repeat all steps leading to~(\ref{eq:totalPsi}). The superscripts
in the equations above indicate the behavior of these functions at
$x \rightarrow - \infty$ corresponding to $1-z=0$. We can deduce this behavior
by comparing~(\ref{eq:phinew}) with~(\ref{eq:lim2}).

A second approach for rewriting~(\ref{eq:totalPsi}) is to start out with the
four independent solutions we found in~(\ref{eq:totalPsi}), use then
{}~(\ref{eq:ident}) and the identity~(\cite{erdelyi})
\be
_{2}F_{1}(a,b,c;1-z) = z^{c-a-b} \ _{2}F_{1}(c-a,c-b,c;1-z)
\ee
to rewrite the four solutions in terms of hypergeometric functions of the
variable $(1-z)$ instead of $z$.

Both approaches yield as result
\be
\Psi_{r}^{s} & = & (A^{s})^{(+ \beta)}  ( \epsilon_{r} \phi_{+}^{(+\beta )}
	{\bf u}_{-}^{s} - 2 \beta \phi_{-}^{(+ \beta)}
		{\bf u}_{+}^{s} ) \nonumber \\
             & + & (A^{s})^{(- \beta)} ( \epsilon_{r} \phi_{+}^{(- \beta)}
	{\bf u}_{-}^{s} + 2 \beta \phi_{-}^{(- \beta)}
		{\bf u}_{+}^{s} )\label{eq:newtotalPsi}
\ee
To impose the boundary conditions for the solution of type $I$ we have to
require that for $x \rightarrow - \infty$ only terms oscillating like
$e^{- 2 \beta x}$ survive. We thus set $(A^{s})^{(+ \beta)} = 0$. The solution
of type $I$ is therefore
\be
(\Psi ^{s}_{r})_{I} = A_{I} ( \epsilon_{r} \phi_{+}^{(- \beta)} {\bf u}_{-}^{s}
+ 2 \beta \phi_{-}^{(- \beta)} {\bf u}_{+}^{s} )
\label{eq:completesolutionI}.
\ee
Continuing now analogously to equations~(\ref{eq:PsitransII}) to~(\ref{eq:RT})
we find the transmitted wave
\be
(\Psi_{r}^{s})_{I}  \stackrel{\mbox{\tiny $x \rightarrow - \infty
$}}{\longrightarrow}
(\Psi _{r}^{s})^{\it trans}_{I} = A_{I} (\epsilon_{r}{\bf u}_{-}^{s}
+ 2 \beta {\bf u}_{+}^{s}) e^{-2\beta x}
\label{eq:PsitransI}
\ee
and the incident and reflected waves for this case
\be
\Psi_{r}^{s} \stackrel{\mbox{\tiny $x \rightarrow + \infty $}}{\longrightarrow}
(\Psi_{r}^{s})^{\it inc}_{I} + (\Psi_{r}^{s})^{\it ref}_{I} \nonumber
\ee
\be
(\Psi_{r}^{s})^{\it inc}_{I} & = & \frac{A_{I} \Gamma (1 - 2\beta)\Gamma
(-2\alpha )}
	           {\Gamma (-\alpha -\beta +\xi)\Gamma (-\alpha -\beta -\xi )}
              \left( \frac {\epsilon_{r}{\bf u}_{-}^{s}}{-\alpha -\beta - \xi}
+
                          \frac {2 \beta {\bf u}_{+}^{s}}{-\alpha -\beta + \xi}
                            \right) e^{-2 \alpha x} \nonumber \\
(\Psi_{r}^{s})^{\it ref}_{I} & = & \frac{A_{I} \Gamma (1- 2\beta)\Gamma (2
\alpha )}
	           {\Gamma (\alpha -\beta +\xi)\Gamma (\alpha -\beta -\xi )}
	  \left( \frac {\epsilon_{r}{\bf u}_{-}^{s}}{\alpha -\beta - \xi} +
                  \frac {2 \beta {\bf u}_{+}^{s}}{\alpha -\beta + \xi}
                            \right) e^{+2 \alpha x } \nonumber \\
\label{eq:inreI}.
\ee
Here we see that we can obtain the reflected wave from the incident wave by
exchanging $\alpha$ with $- \alpha$.

As before we can calculate the currents and find
\be
(j_{I}^{\it inc})_{3} & = & \; \: 8 |A_{I}|^{2}|\alpha | \epsilon_{r}
                         \left| \frac{\Gamma (1 + 2\beta )\Gamma (2\alpha )}
                                     {\Gamma (1 +\alpha +\beta +\xi)\Gamma
(\alpha +\beta -\xi )}
                        \right| ^{2} \nonumber \\
(j_{I}^{\it ref})_{3} & = & \! \! -8 |A_{I}|^{2}|\alpha | \epsilon_{r}
                         \left| \frac{\Gamma (1 + 2\beta )\Gamma (2\alpha )}
                  {\Gamma (1 -\alpha +\beta +\xi)\Gamma ( -\alpha +\beta -\xi
)}
                        \right| ^{2} \nonumber \\
(j_{I}^{\it trans})_{3} & = & \; \:4 |A_{I}|^{2}\epsilon_{r} \epsilon \nonumber
\\
\label{eq:fnalcurrI}
\ee
If we compare these currents to the ones obtained in type $II$
{}~(\ref{eq:fnalcurrII}) we see that we obtain~(\ref{eq:fnalcurrI}) from
{}~(\ref{eq:fnalcurrII}) by exchanging $\alpha$ with $\beta$. Since the
transmission
and reflection coefficients in~(\ref{eq:RT}) are unchanged under exchange of
$\alpha$ with $\beta$ we find that the reflection and transmission coefficients
for a fermion falling into the asymmetric phase or a fermion escaping the
asymmetric phase are the same for energies above the barrier and are given by
{}~(\ref{eq:RT}).

\section{Normalization and Orthogonality}

In this section we want to construct a general solution to the Dirac equation
using the type $I$ solution given in~(\ref{eq:completesolutionI}) and the
type $II$ solution given in~(\ref{eq:completesolutionII}). The general solution
should have the form
\be
\Psi^{s}_{r} = a^{s}_{r} (\Psi^{s}_{r})_{I} + b^{s}_{r} (\Psi^{s}_{r})_{II}
\label{eq:generalsol}
\ee
were $a,b$ are arbitrary coefficients normalized to 1
\be
1 = (a^{s}_{r})^{2} + (b^{s}_{r})^2.
\ee
In~(\ref{eq:generalsol}) $\Psi_{I}$ and $\Psi_{II}$ should represent then
the normal modes of the system as depicted in figure 3.

To obtain such a solution we have to normalize first $\Psi_{I}$ and $\Psi_{II}$
themselves and then assure that the two solutions form an orthogonal set.
It will turn out that they actually are not orthogonal and we finally will
show how to orthogonalize them to obtain a solution of the
form~(\ref{eq:generalsol}).

To normalize the type $I$ and $II$ solutions given
in~(\ref{eq:completesolutionI}) and in~(\ref{eq:completesolutionII})
we first choose an open interval $(+l,-l)$ on the x-axis of figure 1.
We define now the normalization condition for a state $\Psi$ on this
interval as follows
\be
1 &=& lim_{l \rightarrow \infty}~
\int_{-l}^{+l} dx {\overline{\Psi}} \gamma_{0} \Psi \nonumber \\
  &=& lim_{l \rightarrow \infty}~
\int_{-l}^{+l} dx \Psi^{\dagger} \Psi .
\label{eq:condition}
\ee
For $l \rightarrow \infty$ we recover the usual normalization condition from
{}~(\ref{eq:condition}).

To evaluate the integral in equation~(\ref{eq:condition}) we introduce an
additional length scale $\delta$
on $(-l,+l)$ such that $\delta \ll l$. The scale $\delta$ defines an symmetric
interval around $x=0$ such that inside $[- \delta, +\delta]$ the Higgs
potential $g(x)$ changes rapidly from its value close to zero, to its value
close to $2$.
\be
\int_{-l}^{l} dx \Psi^{\dagger} \Psi =
(\int_{-l}^{-\delta} + \int_{-\delta}^{+\delta}
		+ \int_{+\delta}^{+l} )
{}~dx \Psi^{\dagger} \Psi \nonumber
\ee
If the scale $l$ becomes very large we can replace the wavefunction $\Psi$
on the intervals $(-l,-\delta)$ and $(+\delta,+l)$ with its asymptotic limits
at $x \rightarrow \mp \infty$ respectively and obtain so
\be
\int_{-l}^{l} dx \Psi^{\dagger} \Psi = \int_{-l}^{ }
dx \Psi^{\dagger}_{(-\infty)}
	\Psi_{(-\infty)} + \int_{ }^{+l} dx \Psi^{\dagger}_{(+ \infty)}
	\Psi_{(+ \infty)} + {\cal O}(\delta /l).
\label{eq:norm}
\ee
Here $\Psi_{\pm \infty}$ represents the asymptotic limit of the wavefunction
at $x \rightarrow \pm \infty$ respectively and the integrals have to be
evaluated only at the indicated limits $\pm l$. If we let $l \rightarrow
\infty$
the corrections of order ${\cal O} (\delta /l)$ vanishes and expression
{}~(\ref{eq:norm}) becomes exact. The result diverges in this limit linearly
in $l$ analogous to the divergence in the volume of the plane wave
normalization.

To solve equation ~(\ref{eq:norm}) for the constants $A_{I}$ and $A_{II}$
of $\Psi_{I}$ and $\Psi_{II}$ we use their
asymptotic expressions given by~(\ref{eq:PsitransI}), ~(\ref{eq:inreI}),
{}~(\ref{eq:PsitransII}) and ~(\ref{eq:inreII}) respectively. Using the
orthogonality of the spinors $\bf u$ in~(\ref{eq:spinors}) and the fact that
integrals over oscillating functions average to zero $\int dx e^{i k x} = 0$
we obtain after some algebra
\be
|A_{I}|^2 &=& \frac{\alpha T}{4 l \epsilon^2 \beta }~
	\frac{1}{1+R+ \frac{\beta}{\alpha} T} \nonumber \\
|A_{II}|^2 &=& \frac{\beta T}{4 l \epsilon^2 \alpha }~
	\frac{1} {1+R+ \frac{\alpha}{\beta} T}.
\label{eq:constants}
\ee
As before we can obtain $A_{I}$ from $A_{II}$ by exchanging $\alpha$ with
$\beta$.

To prove the orthogonality of $\Psi_{I}$ and $\Psi_{II}$ we need to evaluate
their overlap integral given by the scalar product
\be
I = \int^{+l}_{-l} dx \Psi _{I}^{\dagger} \, \Psi_{II}
\label{eq:overlapp}
\ee
The integral is defined as before on an interval $(-l,+l)$. If the
overlap integral $I$ in~(\ref{eq:overlapp}) is zero then $\Psi_{I}$ and
$\Psi_{II}$ are orthogonal, if it is non--zero we still have to search for
orthogonal solutions.

Proceeding analogously to equations ~(\ref{eq:condition}) to ~(\ref{eq:norm})
we obtain
\be
I= \int^{+l}_{ } dx (\Psi_{I}^{\dagger})_{(+ \infty)} (\Psi_{II})_{(+ \infty)}
     +\int^{ }_{-l} dx (\Psi_{I}^{\dagger})_{(-\infty)} (\Psi_{II})_{(-\infty)}
\label{eq: scalarp}
\ee
Using again the expressions for the asymptotic wave functions we find
\be
I & =& 8l A_{II} A_{I}^{\star}  \epsilon^2 (\beta - \alpha )
\frac{\Gamma (2 \beta ) \Gamma (- 2 \alpha)}{\Gamma ( \beta - \alpha + \xi)
(\beta - \alpha - \xi + 1 ) } \nonumber \\
 & \neq & 0
\label{eq:Iis}
\ee
The overlap integral is therefore non--zero, $\Psi_{I}$ and $\Psi_{II}$ are
not orthogonal. To obtain a solution containing normal modes like it was
envisioned in~(\ref{eq:generalsol}) requires therefore to find a set of
orthogonal states. This can be done using the Schmidt Orthogonalization
{}~(\cite{arfken}).

In the Schmidt Orthogonalization we start out with a set of non--orthogonal
states and construct out of these one by one an orthonormal basis. The set of
non-orthogonal states is in our case $\Psi_{I}$ and $\Psi_{II}$. Let's
choose now $\Psi_{I}$ as our first normalized basis state. The second
orthonormal
basis state corresponding to $\Psi_{I}$ is then obtained through the ansatz
\be
\Psi_{ortho} = N ( a \Psi_{I} + \Psi_{II} ) .
\label{eq:ansatzo}
\ee
Here N and a are constants to be fixed by the requirement that $\Psi_{ortho}$
is normalized to one and is orthogonal to $\Psi_{I}$ respectively.

The orthogonality condition is evaluated by forming the scalar product of
$\Psi_{ortho}$ in~(\ref{eq:ansatzo}) and $\Psi_{I}$
\be
\int dx \Psi_{I}^{\dagger} \Psi_{ortho} &=& N(a\int dx \Psi_{I}^{\dagger}
\Psi_{I} + \int dx \Psi_{I}^{\dagger} \Psi_{II} )  \nonumber \\
& \equiv & 0 \label{eq:ortho}
\ee
The first integral on the right hand side is one since $\Psi_{I}$ is normalized
and we obtain therefore
\be
a=-\int dx \Psi_{I}^{\dagger} \Psi_{II} \equiv -I .
\label{eq:ais}
\ee
The constant $a$ is therefore identical to the overlapp integral $I$ evaluated
in~(\ref{eq:Iis}). It is now straightforward to normalize $\Psi_{ortho}$.
The normalization condition reads
\be
1 = \int dx \Psi_{ortho}^{\dagger} \Psi_{ortho}
\label{eq:overlappo}
\ee
which can be solved for
\be
|N|^2 = \frac{1}{1+|I|^2}
\label{eq:normo}
\ee
We therefore found a general solution of the form~(\ref{eq:generalsol}) with
$\Psi_{II}$ replaced by $\Psi_{ortho}$. Of course, we could have started in
{}~(\ref{eq:ansatzo}) with the orthonormal basis state to $\Psi_{II}$
\be
\Psi_{ortho} ' = N'(a' \Psi_{II} + \Psi_{I}) .
\ee
It is straightforward to see that in this case $a'=-I^{\star}$ and
$|N'|^{2} = |N|^{2}$, so that up to a phase
$\Psi_{ortho}' = \Psi_{ortho}^{\star}$. Again we can construct the general
solution of normal modes, this time replacing $\Psi_{I}$
in~(\ref{eq:generalsol})
with $\Psi_{ortho}^{\star}$.

\section{Conclusion}

In this paper we discussed the motion of fermions under the influence of
electroweak domain walls. We showed that such fermions are well described by
a Dirac equation with an effective mass term. This mass term was obtained
via the Yukawa coupling in the Lagrangian applying the classical mean field
approximation. We substituted for the finite temperature vev of the Higgs
field the classical solution to the equations of motion of a Higgs field
in a finite temperature effective potential.

We investigated the Dirac equation analytically and found the wave function
of the fermion. Transmission and reflection coefficients were derived and
found to be the same for both fermions and antifermions.

\section*{Acknowledgment}

We would like to thank R.~Venugopalan for helpful discussions.
Two of us, A. A. and J. J. M., would also like to thank
R. Madden and R. Rodriguez for their useful comments.
L. McLerran wants to thank Mikhail Shaposhnikov for
crucial comments early on in this work.
L. McLerran wishes to also acknowledge the Aspen Center for
Physics where part of this work was completed.
This work was supported by the U. S. Department
of Energy under grants DOE/DE--AC02--83ER40105 and DOE/DE--FG02--87ER40328,
by the Gesellschaft f\"ur
Schwerionenforschung mbH under their program in support of university
research and by the DGAPA/UNAM/M\'{e}xico.

\newpage

\section*{Appendix A}
\renewcommand{\theequation}{\roman{equation}}
\setcounter{equation}{0}

Here, we would like to show that $\varphi$'s defined as
\be
\varphi _{\pm} = [\mp \frac{d}{dx} + \xi g(x)] \phi _{\pm}
\ee
in equation~(\ref{eq:epsindcomp}) are proportional to the $\phi$ 's given by
equation~(\ref{eq:solution phi}) where $g(x) = 1 + \tanh x $. We will prove it
for one case, show that it can be done for the others and write the
specific relations for all cases.

Let's start by looking at one case, for example,
\be
\varphi _{+}^{(-\alpha)} = [- \frac{d}{dx} + \xi g(x)] \,\phi _{+}^{(-\alpha)}
\ee
where
\be
\phi _{+}^{(-\alpha)} = z^{\alpha} (1-z)^{\beta}\ _{2}F_{1}(\alpha + \beta -
\xi + 1, \alpha + \beta + \xi,
2 \alpha + 1, z)
\ee
and $z= \frac{1}{2}(1 - \tanh x )$. Then it follows that
\be
\ [- \frac{d}{dx} + \xi g(x)] = 2 (1 -z)[ z \frac{d}{dz} + \xi ]
\ee
so we can rewrite
$\varphi _{+}^{(-\alpha)} $ as
\be
\varphi _{+}^{(-\alpha)} =2 (1 -z)[ z \frac{d}{dz} + \xi ] \phi
_{+}^{(-\alpha)}.
\ee
Let's consider
\be
\frac{d}{dz} [z^{\xi} \phi _{+}^{(-\alpha)}] = \xi z^{\xi - 1} \phi
_{+}^{(-\alpha)} +
z^{\xi }\frac{d}{dz}\phi _{+}^{(-\alpha)}                \nonumber
\ee
Multiplying both sides of this equation
 by $2(1- z) z^{-\xi + 1}$,\,  we obtain
\be
2(1- z) z^{-\xi + 1} \frac{d}{dz}[ z^{\xi}\, \phi _{+}^{(-\alpha)}] =
2(1-z)[z \frac{d}{dz} + \xi ] \phi _{+}^{(-\alpha)}.    \nonumber \\
\ee
Comparing this with equation~(\ref{eq:solution phi})
and using the explicit form for $\phi$, we have
\be
\varphi _{+}^{(-\alpha)}&=& 2(1- z) z^{-\xi +1}\frac{d}{dz}[z^{\xi +
\alpha}(1-z)^{\beta}\ _{2}F_{1}(\alpha +
                       \beta - \xi + 1, \alpha + \beta + \xi, 2\alpha + 1, z)]
\nonumber \\
       \,       &=& 2(1-z)z^{\alpha + 1}[\frac{d}{dz} (1-z)^{\beta}]\
_{2}F_{1}(\alpha +
                       \beta - \xi + 1, \alpha + \beta + \xi, 2\alpha + 1, z)
\nonumber \\
         \,     &+& 2(1-z)^{\beta +1} \,z^{-\xi +1} \frac{d}{dz}[z^{\xi +
\alpha }\,\ _{2}F_{1}(\alpha +
                       \beta - \xi + 1, \alpha + \beta + \xi, 2\alpha + 1, z)]
\nonumber
\ee
We now add and subtract $\beta$ to the exponent of $z$
inside the bracket in the second term above
and differentiate to get
\be
         \, & \frac{d}{dz} & [z^{\alpha + \xi} \ _{2}F_{1}(\alpha + \beta - \xi
+ 1,
                                 \alpha + \beta + \xi, 2\alpha + 1, z)]=
\nonumber \\
 z^{-\beta} & \frac{d}{dz} & [z^{\alpha +\beta +\xi} \ _{2}F_{1}(\alpha +
                                 \beta - \xi + 1, \alpha + \beta + \xi, 2\alpha
+ 1, z)]
									 \nonumber \\
         \, &      -       & \beta z^{\alpha + \xi -1} \
				     _{2}F_{1}(\alpha + \beta - \xi + 1, \alpha +
					       \beta + \xi, 2\alpha + 1, z)  \nonumber
\label{eq:der}.
\ee
Now we use the identity~(\cite{erdelyi})
\be
\frac{d}{dz}[ z^{b} \ _{2}F_{1}(a,b,c,z)]= bz^{b-1}\ _{2}F_{1}(a,b+1,c,z)
\nonumber
\ee
and putting everything together, we find
\be
\varphi _{+}^{(-\alpha)} &=& -2\beta z^{\alpha}(1-z)^{\beta}\ _{2}F_{1}(\alpha
+
                       \beta - \xi + 1, \alpha + \beta + \xi, 2\alpha + 1, z)
\nonumber \\
             \,  &+& 2(\alpha + \beta + \xi)z^{\alpha}(1-z)^{\beta + 1}\
_{2}F_{1}(\alpha +
                       \beta - \xi + 1, \alpha + \beta + \xi + 1, 2\alpha + 1,
z) \nonumber
\ee
In the second term of this equation we can use the identity~(\cite{erdelyi})
\be
\ _{2}F_{1}(a,b,c,z)=\frac{c-a}{(b-a)(1-z)}\ _{2}F_{1}(a-1,b,c,z) -
\frac{c-b}{(b-a)(1-z)}
\ _{2}F_{1}(a,b-1,c,z)    \nonumber
\ee
and then combine the terms such that
\be
\varphi _{+}^{(-\alpha)} & = & [-2\beta -\frac{1}{\xi}[\alpha ^{2} -(\beta +
\xi )^{2}]]z^{\alpha}(1-z)^{\beta}
                        \ _{2}F_{1}(\alpha + \beta -\xi +1, \alpha + \beta
+\xi, 2\alpha +1,z)
                            \nonumber \\
             \,  & + & \frac{1}{\xi}[(\alpha + \xi)^{2} - \beta
^{2}]z^{\alpha}(1-z)^{\beta}
                        \ _{2}F_{1}(\alpha + \beta -\xi, \alpha + \beta +\xi
+1, 2\alpha +1, z)
                         \nonumber
\ee
Using the fact that $\alpha ^{2} = \beta ^{2} + \xi ^{2}$, we see that the
coefficient of the
 first term is identically zero and that
\be
\frac{1}{\xi}[(\alpha + \xi)^{2} - \beta ^{2}]= 2(\xi + \alpha)  \nonumber
\ee
Finally comparing to equation~(\ref{eq:solution phi})
\be
\phi ^{(-\alpha)}_{-} = z^{\alpha}(1-z)^{\beta}\ _{2}F_{1}(\alpha + \beta -\xi,
\alpha + \beta + \xi + 1,
2\alpha +1,z)   \nonumber
\ee
we see that
\be
\varphi ^{(-\alpha)}_{+} = 2(\xi + \alpha) \phi ^{(-\alpha)}_{-}
\ee

Similarly we can show that
\be
\varphi ^{(-\alpha)}_{-} = 2(\xi - \alpha) \phi ^{(-\alpha)}_{+} \nonumber \\
\varphi ^{(+\alpha)}_{+} = 2(\xi - \alpha) \phi ^{(+\alpha)}_{-} \nonumber \\
\varphi ^{(+\alpha)}_{-} = 2(\xi + \alpha) \phi ^{(+\alpha)}_{+}
\ee
For the case when $\varphi $'s and $\phi $'s are expanded around $(1- z)$
instead of $z$, we can
apply the same method to derive similar relationships. They are
\be
\varphi _{+}^{(-\beta)} = 2\beta \phi _{-}^{(-\beta)} \nonumber
\ee
\be
\varphi _{+}^{(+\beta)} = -2\beta \phi _{-}^{(+\beta)}
\ee
These relations are used in the calculation of the fermionic wave function
to eliminate the linearly dependent solution.

\newpage

\section*{Figure captions}

Figure 1: The bubble profile. $x=-\infty $ corresponds to the bubble's
outside. The domain wall is the region where the vev of the Higgs field
$ g(x) $ changes rapidly.\\
Figure 2: The dimensionless effective potential for the Higgs field
$V(g, \zeta)$ in units of $\delta T ( \frac{\delta T}{\lambda})^{3}$
for different values of $ \zeta = 0, 2, 1.75, 2.25, 2.75$.\\
Figure 3: Scetch of the asymptotic behavior of the solutions  of type $I$
and $II$.\\
Figure 4: Transmission and reflection coefficients for $\xi = 0.6,4$ which
correspond to a ratio of fermion to Higgs mass of 0.3 and 2 respectively.
The reduced energy parameter is $y=\frac{\epsilon}{2 \xi}$, so that the
top of the barrier is at $y=1$.

\end{document}